 \documentclass[
 reprint,
 amsmath,amssymb,
 aps,
 pra,
floatfix,
]{revtex4-2}

\usepackage{graphicx}
\usepackage[export]{adjustbox} 

\usepackage{dcolumn}
\usepackage{bm}
\usepackage{booktabs}
\usepackage{tikz}
\usetikzlibrary{quantikz}
\usepackage{hyperref}
\usepackage{subcaption} 
\usepackage{orcidlink}
\begin{document}


\title{Implementation of classical client universal blind quantum computation with 8-state RSP in current architecture}

\author{Aman Gupta \orcidlink{0000-0003-2918-1952}}
 \email{aman.ag220@gmail.com}
\author{Daniel Prasanth}%
\author{Venkat Chandra Gunja}

\date{\today}

\begin{abstract}
The future of quantum computing architecture is most likely the one in which a large number of clients are either fully classical or have a very limited quantum capability while a very small number of servers having the capability to perform quantum computations and most quantum computational tasks are delegated to these quantum servers. In this architecture, it becomes very crucial that a classical/semi-classical client is able to keep the delegated data/ computation secure against eavesdroppers as well as the server itself, known as the blindness feature. 
 In 2009, A. Broadbent et. al proposed a universal blind quantum computation (UBQC) protocol based on measurement-based quantum computation (MBQC) that enables a semi-classical client to delegate universal quantum computation to a quantum server, interactively and fetch the results while the computation itself remains blind to the server. In this work, we propose an implementation (with examples) of UBQC in the current quantum computing architecture, a fully classical client, a quantum server (\textit{IBM Quantum}) and the computation does not proceed interactively (projective measurement basis is not decided by previous measurement outcome). We combined UBQC with the 8-state remote state preparation (RSP) protocol, to blindly prepare the initial cluster state, which is an initial resource state in UBQC protocol, to allow a completely classical client to perform delegated blind quantum computation. Such an implementation has already been shown to be secure in a game-based security setting, which is the weakest security model.   
\end{abstract}

\maketitle


\section{Introduction}
Quantum computing is a fast-developing field that offers solutions to problems that classical computers couldn't solve in polynomial time, especially the bounded quadratic polynomial (BQP) problems, the holy grail of computing. As technology advances, the setup of quantum computation is and will be a client-server model, where most users rely on fully classical or semi-classical devices and quantum computations are performed on specialized servers. This arrangement, however, raises concerns about security on how users can ensure their computations stay private, not just from outside threats but also from the quantum server itself.

Blind quantum computation (BQC) is a technique that addresses this issue by enabling users to perform their quantum computations without revealing sensitive information. A key milestone in this field was the UBQC protocol, known as BFK protocol, introduced by A. Broadbent, et. al in 2009\cite{Broadbent_2009}. UBQC uses a measurement-based approach, where a client encodes their computation into a prepared quantum state and performs adaptive measurements to run the algorithm. This allows a semi-classical client to interact securely with a quantum server without exposing any details of their computation.

However, implementing UBQC on modern quantum platforms like IBM Quantum is tricky, as many of these systems don’t support the real-time interaction required for adaptive measurements. To overcome this, we modified UBQC by eliminating the need for interactive measurements and replacing them with pre-defined measurement bases. We also incorporated the 8-state Remote State Preparation (RSP) protocol, enabling a fully classical client to prepare the necessary quantum resource state and post-computation filtering.

Our approach makes it possible for classical clients to perform secure computations on quantum servers. This brings us closer to making UBQC practical for real-world applications.
\section{Background}
\subsection{UBQC}
 Raussendorf and Briegel \cite{Raussendorf_2003}, proposed MBQC, a method alternative and completely equivalent to the circuit model of quantum computation, to perform universal quantum computation by only doing the $X$ and $Z$ basis measurements on a highly entangled $2D$ resource state known as Cluster State. The BFK protocol uses the above universal MBQC method to provide a universal blind quantum computation scheme. This was achieved by strengthening the requirement of initial states of the cluster state to be in an arbitrary $\ket{+_{\theta_{i,j}}}$ states (known only to the client) and relaxing the measurement requirement to be $Z$ basis and any projective measurement in the orthogonal $X-Y$ plane, the server entangles the state received by the client forming the intermediate cluster state with neighboring entanglements and then measurement being performed by the server based on instructions by the client to complete the quantum computation (see Fig. \ref{M2UBQC}). The detailed steps in the UBQC protocol are given below: 
\begin{enumerate}
    \item \textbf{State Preparation} \label{state+prep_step}
    \begin{itemize}
        \item A client (Alice) with the state preparation ability prepares quantum states in
        \begin{equation}
        \label{ini_state}
        \ket{+_{\theta_{i,j}}} = \frac{1}{\sqrt{2}}(\ket{0}+e^{i\theta_{i,j}})
        \end{equation}
        $\forall \theta_{i,j} \in \{0.\frac{\pi}{4},1.\frac{\pi}{4},2.\frac{\pi}{4},...7.\frac{\pi}{4}\}$ and sends it to a quantum server (Bob) via quantum channel.
        \item The server upon receiving the qubits entangles them to create a cluster state. 
    \end{itemize}
    \item \textbf{Projective measurement}
    \begin{itemize}
        \item Alice then computes the projective measurement angle as \begin{align}
        \label{eq_totalangle}
        \delta_{i,j} = \phi_{i,j}'+\theta_{i,j}+\pi r_{i,j}\nonumber \\ \text{ where, }\phi_{i,j}' = (-1)^{s^X_{x,y}}\phi_{x,y}+s^Z_{x,y}\pi \end{align}.
        \item Alice sends this angle information via a classical channel to Bob.
        \item Based on the information received Bob applies the measurement operator $M(\delta_{i,j})$ on the \{$i,j$\}$^{th}$ qubit in the basis $\{\ket{+_{\delta_{i,j}}},\ket{-_{\delta_{i,j}}}\}$.
        \item Bob sends the measurement result $s_{i,j}$ to Alice.
    \end{itemize}
    \item \textbf{Correction}\\
    If $r_{i,j} =1$, Alice flips the result $s_{i,j}$, else she accepts the result as is.
    \end{enumerate}
There also exist some variations of this protocol where the client performs single-qubit projective measurements on a qubit from the cluster state sent by the server\cite{PhysRevA.87.050301}.

\begin{figure*}[htb] 
    \centering

    \begin{subfigure}[b]{0.45\textwidth} 
        \centering
        \includegraphics[width=\linewidth]{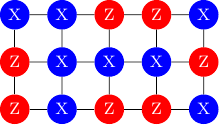} 
        \caption{An example MBQC cluster state with $X$ and $Z$ basis measurements, the initial state of all the nodes in this cluster are $\ket{+} = \frac{1}{\sqrt{2}}(\ket{0}+\ket{1})$}.
        \label{MBQC_fig}
    \end{subfigure}
    \hfill
    \begin{subfigure}[b]{0.45\textwidth}
        \centering
        \includegraphics[width=\linewidth]{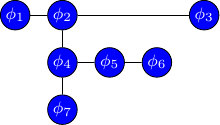} 
        \caption{UBQC version of same cluster state with only projective measurements in $X-Y$ plane, the initial state of each node in this cluster is given in Eq.\ref{ini_state}}.
        \label{UBQC_fig}
    \end{subfigure}

    \caption{\ref{MBQC_fig} shows an MBQC model based 2D cluster state computation and \ref{UBQC_fig} is an equivalent cluster state in UBQC model.}
    \label{M2UBQC}
\end{figure*}

\subsection{8-State Remote State Preparation}
\label{section_rsp}
To prepare a quantum state of the form $\ket{\psi} = \frac{1}{2}[\ket{0}+e^{i\theta}\ket{1}]$, where $\theta \in  \{0.\frac{\pi}{4},1.\frac{\pi}{4},2.\frac{\pi}{4},...7.\frac{\pi}{4}\}$, by a fully classical client, which is unknown even to the quantum server at its end, 
Alexandru Cojocaru et. al. proposed a quantum oracle, constructed using a family of trapdoor one-way functions that are two-regular and collision resistant even against quantum adversary\cite{cryptography5010003}, named QFactory protocol. 8-state because this algorithm is able to prepare a secret state from a set of eight possible states. We used the exact recipe of 8-state RSP with a very simple and particular instance of such a trapdoor function, given below, to implement the non-interactive classical client 8-state RSP protocol to create one quantum state of the given form.
\begin{itemize}
    \item \textbf{Step 1: Key Generation Algorithm}
    \begin{itemize}
        \item Client samples uniformly at random a public bit string $\alpha_i \in \{0,1\}^3$ and a private trapdoor information, $t_k= (d_0,e) \in \{0,1\}^2$.
        \item Client then constructs a public key pair, $(A,B)$ as a function of the trapdoor in the following way:\\
        $A_{e+1,e+1} = A_{2-e,3}=B_{3,3} =1$\\
        $B_{2-e,2-e} = d_0$; all other elements of $A,B$ matrices are 0.
        \item Example: If $t_k = (1,0); d = 1, e=0$;\\
        $A_{1,1} = A_{2,3} = B_{3,3} = 1$, $B_{2,2} =1$\\
        $\implies A = \begin{bmatrix}
            1 & 0 & 0\\
            0 & 0& 1\\
            0 & 0&0 \end{bmatrix}$ and $B = \begin{bmatrix}
                0 & 0& 0\\
                0&1&0\\
                0&0&1
            \end{bmatrix}$
    \end{itemize}
    \textbf{Step 2: Constructing the quantum trapdoor oracle}\\
    For the trapdoor function $f_{A,B} : \{0,1\}^3 \rightarrow \{0,1\}^2$ and public key pair $(A,B) \in \{0,1\}^{3\times3}$. The client instructs the quantum server to:  \begin{itemize}
    \item create a quantum gate (say, Oracle) with five qubits, the first 3 qubits as input (control) qubits, and the remaining two qubits as output (target) qubits. 
    \item implement a toffoli gate (or CNOT gate, if $i==j$) between $i^{th}$ input qubit, $j^{th}$ input qubit and first output qubit whenever $A_{i,j} = 1$.
    \item Similarly, implement a toffoli gate (or CNOT gate, if $i==j$) between $i^{th}$ input qubit, $j^{th}$ input qubit and second output qubit whenever $B_{i,j} = 1$.\\
    \end{itemize}
    \item \textbf{Step 3: RSP circuit}
\begin{itemize}
    \item Now the client instructs the server to create a quantum circuit with five quantum bits and four classical bits ($y,b = y_1y_2b_1b_2$).
    \item The first two qubits represent control qubits that compute the output image of the trapdoor oracle, while the last two target qubit compute the parameter $b$ that is used to compute the $\theta$ in later stage. The third control qubit is where the desired state is generated.
    \item The client instructs the server to apply the Hadamard gate on all three control qubits and the oracle is composed.
    \item A squeezing operation is then added to the circuit, which is just a projective measurement in $\ket{\pm_{\alpha\pi/4}}$, the result of which is stored in classical bits $y = y_1y_2$
    \item Finally, the last two control qubits are measured and the results are stored in classical bits $b = b_1b_2$.
\end{itemize}
    \item \textbf{Step 4: Inversion function to get-preimages}
    \begin{itemize}
        \item Given the function's ($f$) image $y = (y_1,y_2) \in \{0,1\}^2$, its two pre-images are given by $x$ and $x'$ as:\\
        $x_{2-e} = 0, x_{1+e} = y_1, x_3=y_2$\\
        $x'_{2-e} =1, x'_{1+e} = y_1 \oplus y_2 \oplus d_0, x_3'=y_2 \oplus d_0$
        \item We compute the two pre-images $x = (x_1,x_2x_3) \in \{0,1\}^3$ and $x' = (x'_1,x'_2,x'_3) \in \{0,1\}^3$ with the equations given above.
        \item Example: for $t_k = (1,0)$ and $y = (1,1)$:\\
        $x_2 = 0,x_1 =1, x_2 =1, x = 101$\\
        $x'_2 =1, x'_1 = 1 \oplus 1 \oplus 1 = 1, x'_3 = 1 \oplus 1 =0, x' = 110$.
        \item If $x_3 == x'_3$, then State preparation algorithm failed, use a different trapdoor sample. 
    \end{itemize}
    \item \textbf{Step 5: Client calculates $\theta$.}
    \begin{itemize}
        \item The server measures all the qubits except the last input qubit and returns the bitstring $b$ to the client.
        \item  The client calculates $\theta$ from the pre-images ($x,x'$), $\alpha$ and measurement outcome $b = (b_1b_2) | b_j \in \{0,1\}$ as:\\
$\theta = \frac{\pi}{4}(-1)^{x_n}\big(\sum_{i=1}^4(x_i-x_i')(4b_i+\alpha_i)\big) \text{ mod 8}.$
    \item  If the protocol is run honestly by the server, the state prepared at the server's end is $\ket{+_\theta} = \frac{1}{\sqrt{2}}(\ket{0}+e^{i\theta}\ket{1})$, where $\theta$ is the value computed by the client.
    \end{itemize}
\end{itemize}

\section{Our method}
Almost every literature in the direction that is focused towards classical client blind quantum computation \cite{10.1007/978-3-030-34578-5_22,cryptography5010003,gheorghiu2019computationallysecurecomposableremotestate, 10.1007/978-3-030-64834-3_23,morimae2019impossibilityperfectlysecureonerounddelegated,10.1007/978-3-030-90459-3_1,Xu_2022} present the compossible security arguments for any kind of classical/semi-classical client delegated blind computation, be it the UBQC type (BFK protocol) or quantum homomorphic encryption type. We, in this work, relied on the game-based security proof (which is the weakest security model) of the QFactory protocol composed with UBQC \cite{10.1007/978-3-030-64834-3_23} and discussed the step-by-step implementation of such a composed blind quantum computing protocol in the current architecture. We used the Qiskit v0.43.0 platform as our state-of-the-art features accessible to a classical client and superconducting architecture-based quantum simulator (QASM simulator). 
We demonstrated the correctness of our method by executing some examples with our proposed methodology, see Sec.\ref{examples}. The examples were delegated to run on an IBM quantum simulator
. Given below are the steps that allow any quantum circuit described in the circuit model or in the cluster state model to be executed in a blinded (encrypted) manner:
\begin{itemize}
    \item \textbf{Step 0: Conversion} \\Since we are using the UBQC protocol, we need to convert a quantum circuit described in the circuit model to an MBQC model using the equivalence relations given in Appendix \ref{cir2MBQC}, we further use the gate identities (see Appendix \ref{gate_identity}) to simplify the gate complexity of the circuit. We now have a cluster state with only the projective measurements in the $X-Y$ plane. In addition, we also have to instruct a quantum server to create a cluster state, (as in Fig. \ref{MBQC_fig}), therefore it is more advantageous to use a brickwork state (as in Fig.\ref{UBQC_fig}), requiring less entanglement operations\cite{Broadbent2010}.
    \item \label{Step 1} \textbf{Step 1: Randomization of projective measurement}\\ One can start from this step if they already have a quantum computation in the brickwork state model. If we remember in UBQC the blindness comes from the fact that the initial state is in the form of $\ket{+_{\theta_{i,j}}}$ where $\theta_{i,j} =k.\frac{\pi}{4};$ $k \in \{0,1,2,3,4,5,6,7\}$, is only known to the client and a randomization parameter $r_{i,j}$. Therefore we introduce additional random rotations $\theta_{i,j}$ in both the projective measurement and initial state, As opposed to the original UBQC protocol, which uses another randomization parameter $r_{i,j}$ and the correction is applied to the next projective measurement parameter itself by $\phi'$ (see Eq. \ref{eq_totalangle}), we cannot do both of these things as current architecture doesn't allow a classical client to perform interactive computation, we therefore, push the complete correction calculus to the end of the circuit (to output qubit), as seen in Fig. \ref{1Dcircuit_blind}, and we don't use $r_{i,j}$. We show that both methods of projective measurement correction are completely equivalent and that the introduction of $r_{i,j}$ doesn't work in Appendix\ref{blindness_appendix}. The resulting change in projective measurement angle is  $\delta_{i,j} = \phi_{i,j}-\theta_{i,j}$, while this was $\delta_{i,j} = \phi'_{i,j}+\theta_{i,j}+r_{i,j}\pi$ in the original UBQC proposal (see Sec. \ref{state+prep_step}). 
    \begin{figure*}[htb] 
    \centering

    \begin{subfigure}[b]{0.45\textwidth} 
        \centering
        \includegraphics[scale=0.9]{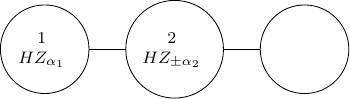} 
        \caption{A 1D cluster state implementing single qubit gate operation}
        \label{1Dcluster_fig}
    \end{subfigure}
    \hfill
    \begin{subfigure}[b]{0.45\textwidth}
        \centering
        \includegraphics[width=\linewidth]{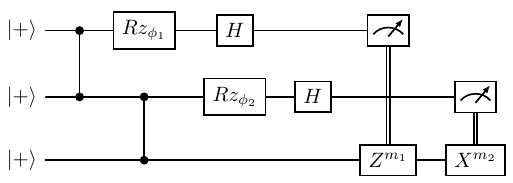} 
        \caption{The equivalent circuit implementation of the given 1D cluster. The $\pm \phi$ in first is converted to Pauli Corrections\cite{NIELSEN2006147}.}
        \label{1Dcircuit}
    \end{subfigure}
  \begin{subfigure}[b]{0.45\textwidth} 
        \centering
        \includegraphics[width=\linewidth]{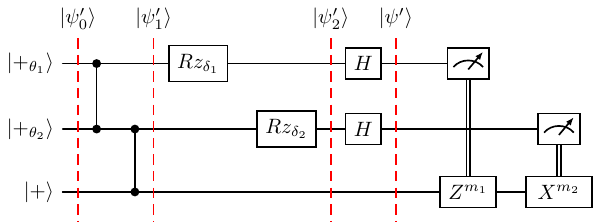} 
        \caption{Blinded UBQC circuit implementing the 1D cluster. Here $\delta_i = \phi_{i}-\theta_{i}$. This is completely equivalent to Fig.\ref{1Dcircuit} because $Rz$ operation commutes with $Controlled-Z$ gate.}
        \label{1Dcircuit_blind}
    \end{subfigure}
    \caption{Information processing on single qubit through projective measurements, the first qubit is the input qubit, the intermediate qubits perform gate operations and propagate information through One-qubit teleportation operation and the last qubit is the output qubit}
    \label{method_step2}
\end{figure*}
    \item \textbf{Step 2: QFactory implementation}\\ Since our client is a completely classical the initial state $\ket{+_{\theta_{i,j}}}$ is also created at server's end using classical instructions. To ensure that the initial state remains unknown to the server we used a specific instance from a family of LWE functions, as given in Sec. \ref{section_rsp}. We implemented the QFactory as our state preparation orcale.
    \item \textbf{Step 3: Composition}\\ We composed both QFactory state preparation oracle with the UBQC oracle (given in step 1), remember that we could do this under the game-based security model, and send the complete classical instruction of this circuit to the server. An example composed circuit is shown in Fig. \ref{composed_cir}.

    \begin{figure*}[htb] 
    \centering
        \includegraphics[width=\linewidth]{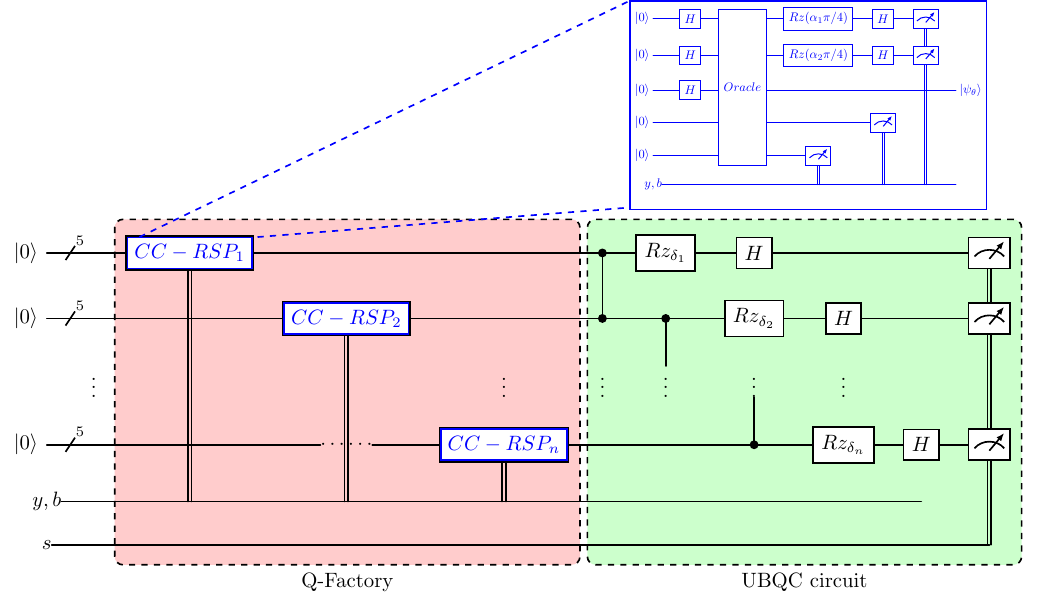} 
        \caption{An example MBQC cluster state with $X$ and $Z$ basis measurements, the initial state of all the nodes in this cluster are $\ket{+} = \frac{1}{\sqrt{2}}(\ket{0}+\ket{1})$}.
    \label{composed_cir}
\end{figure*}

    \item \textbf{Step 4: Filtering}
    \\ Here we assume that all the outputs are classical, (this is a justified assumption as the client is fully classical and all that can be fetched from the server is the classical output). The problem with the QFactory oracle is that it doesn't prepare a pre-determined state rather a quantum state prepared is determined based on the classical results of the oracle execution, therefore, we need to filter only the executions using the classical sub-string, $\prod_n (y_1y_2b_1b_2)_n$,  obtained from measurement result that correspond to the correct $\prod_n \theta_{i,j}$, $\forall n$, where $n$ is the number of qubits in the brickwork state (excluding the output qubit).
\end{itemize}
It can be observed from Step 4, that we need eight additional executions just to create one quantum state, this way we drop the probability of correct success probability of single execution as $1/2^{4n}$, where $n$ is the number of qubits in the blinded circuit, this is an exponential cost in the circuit execution compared to circuit model. One might think that with the new features of mid-circuit measurements provided for classical client one can avoid this exponential execution cost however, in this scenario the trade-off is the blindness itself.

\subsection{Results}
\label{examples}
\begin{figure*}[htb] 
    \centering

    \begin{subfigure}[b]{0.23\textwidth} 
        \centering
        \includegraphics[scale=0.5]{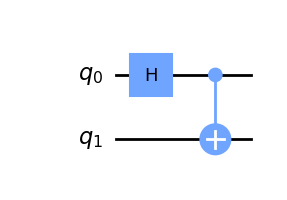} 
        \caption{Quantum circuit for bell state preparation of the first kind, $\ket{\phi^+} = \frac{1}{\sqrt{2}}[\ket{00}+\ket{11}]$}
        \label{bell_ub}
    \end{subfigure}
    \hfill
    \begin{subfigure}[b]{0.74\textwidth}
        \centering
        \includegraphics[width = 14cm, height = 3cm]{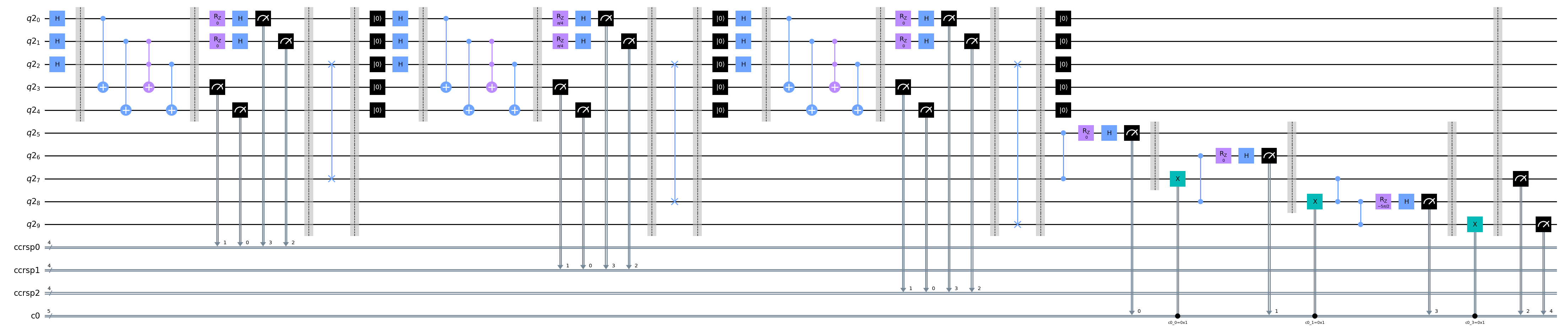} 
        \caption{The equivalent blinded quantum circuit using the proposed composed RSP and UBQC method. Notice that here we have additionally used SWAP gates to reduce the number of qubits used of state preparation.}
        \label{bell_b}
    \end{subfigure}
    \caption{This is an example implementation of delegated private bell state preparation using the given methodology. The Fig.\ref{bell_b} is what was sent to the IBMQ server.}
    \label{bell_example}
\end{figure*}

\begin{figure*}[htb] 
    \centering

    \begin{subfigure}[b]{0.23\textwidth} 
        \centering
        \includegraphics[scale=0.5]{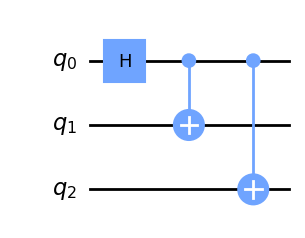} 
        \caption{Quantum circuit for GHZ state preparation, $\ket{GHZ} = \frac{1}{\sqrt{2}}[\ket{000}+\ket{111}]$}
        \label{ghz_cir}
    \end{subfigure}
    \hfill
    \begin{subfigure}[b]{0.74\textwidth}
        \centering
        \includegraphics[width = 11cm, height = 5cm]{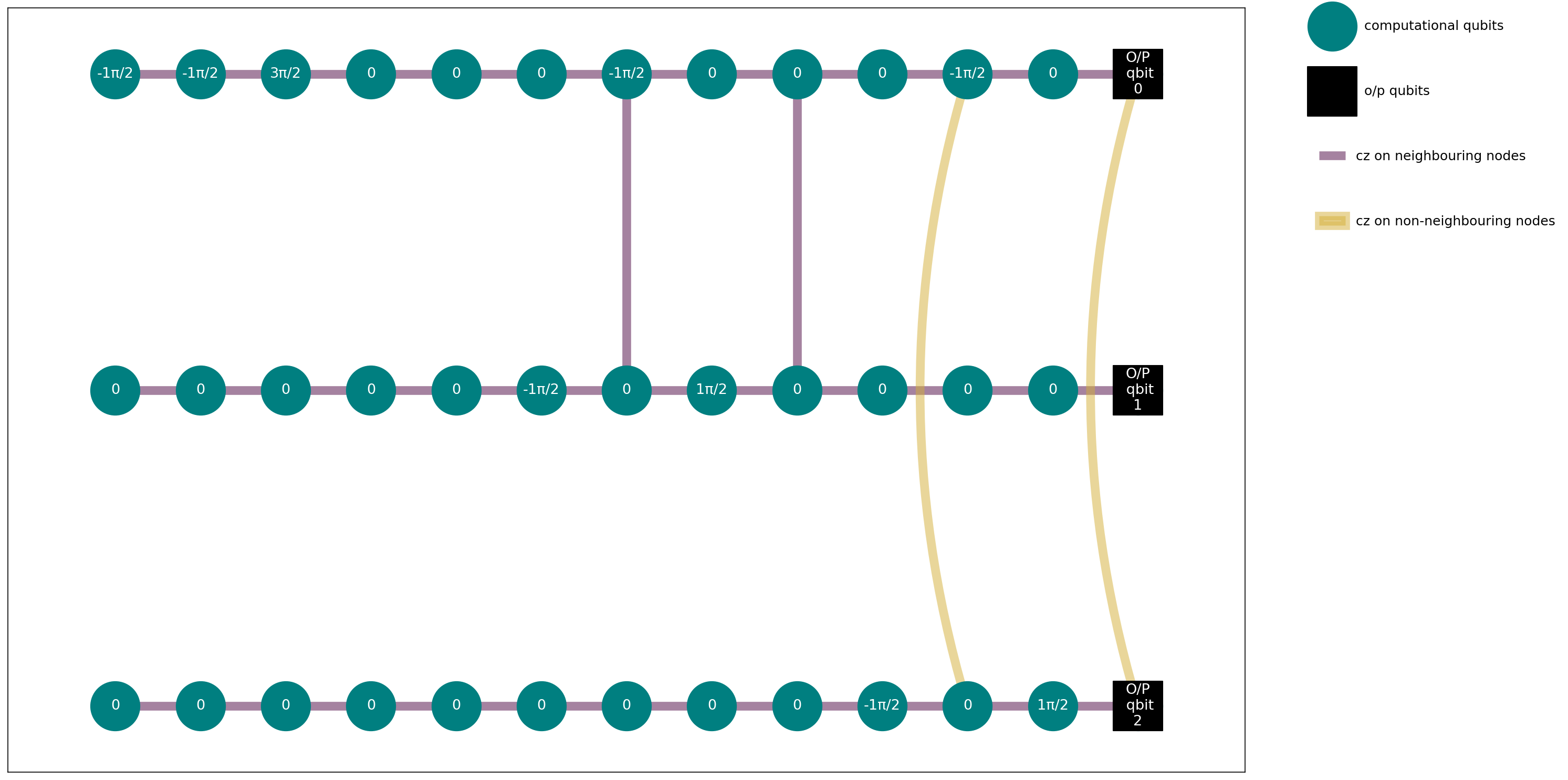} 
        \caption{The unblinded brickwork state for the same GHZ state}
        \label{ghz_ub}
    \end{subfigure}\\
    \begin{subfigure}[b]{0.6\textwidth}
        \includegraphics[width = 11cm, height = 5cm]{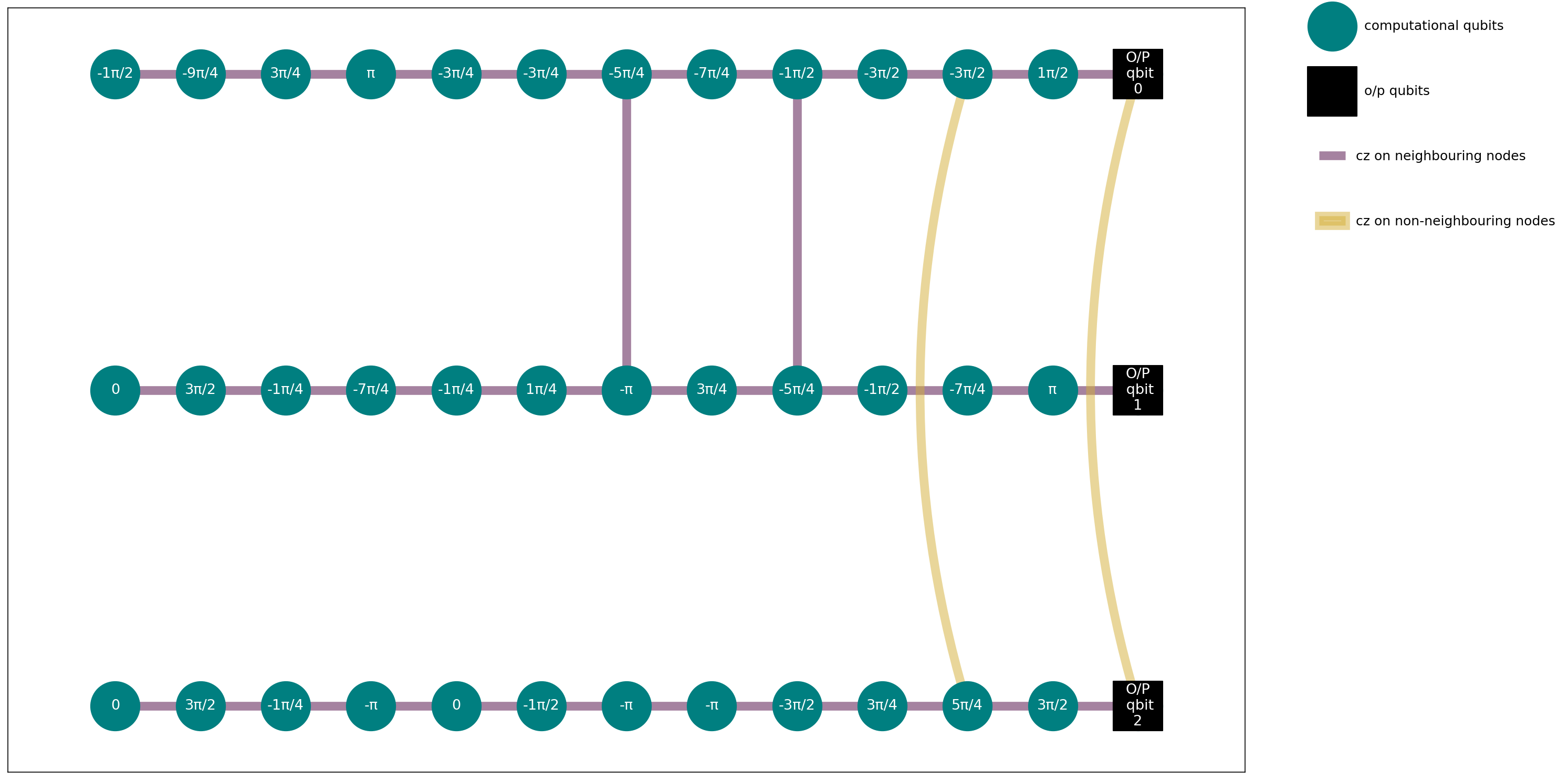} 
        \caption{The blinded brickwork state for the GHZ state preparation circuit.}
        \label{ghz_b}
    \end{subfigure}
    \hfill
    \begin{subfigure}[b]{0.35\textwidth}
        \centering
        \includegraphics[width =\linewidth]{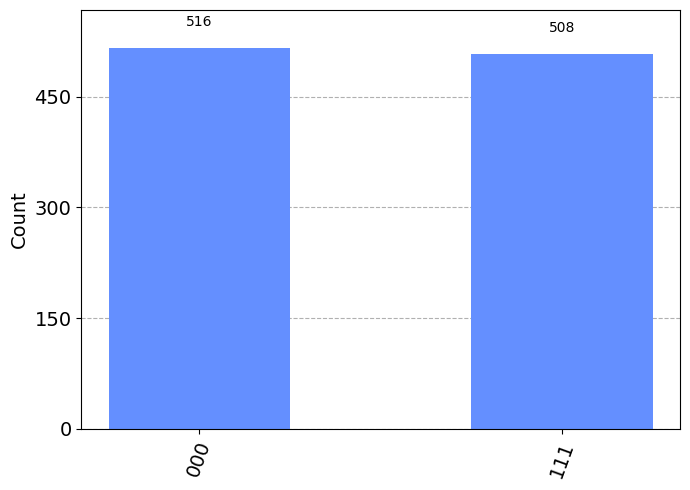} 
        \caption{Output result fetched from the server corresponding to the output qubits after filtering}
        \label{ghz_result}
    \end{subfigure}
    \caption{This is an example implementation of delegated private GHZ state preparation using the given methodology. The quantum circuit corresponding to Fig.\ref{ghz_b} was sent to the IBMQ server.}
    \label{GHZ}
\end{figure*}

\begin{figure*}[htb] 
    \centering

    \begin{subfigure}[b]{0.25\textwidth} 
        \centering
        \includegraphics[width = \linewidth]{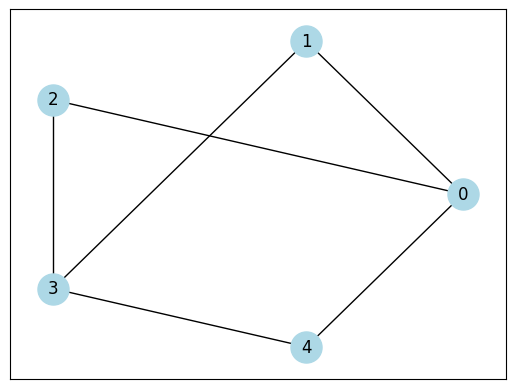} 
        \caption{An example weighted graph with five nodes for the Maximum Weighted Independent set Problem (MWIP).}
        \label{graph}
    \end{subfigure}
    \hfill
    \begin{subfigure}[b]{0.74\textwidth}
        \centering
        \includegraphics[width = 11cm, height = 6.5cm]{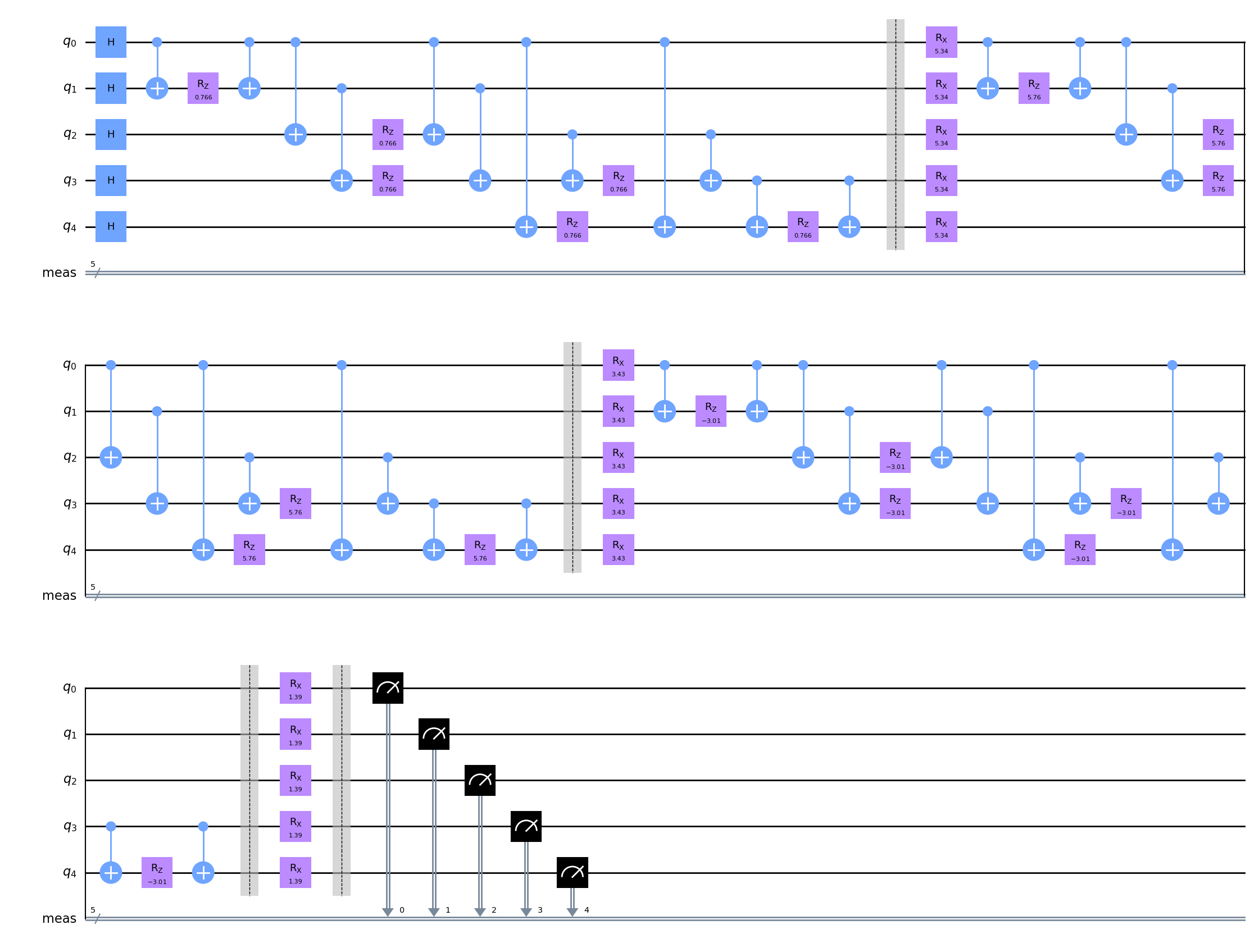} 
        \caption{Quantum circuit corresponding to the Hamiltonian of the example problem.}
        \label{mwip_circ}
    \end{subfigure}\\
    \begin{subfigure}[b]{0.6\textwidth}
        \includegraphics[width = 11cm, height = 6.8cm]{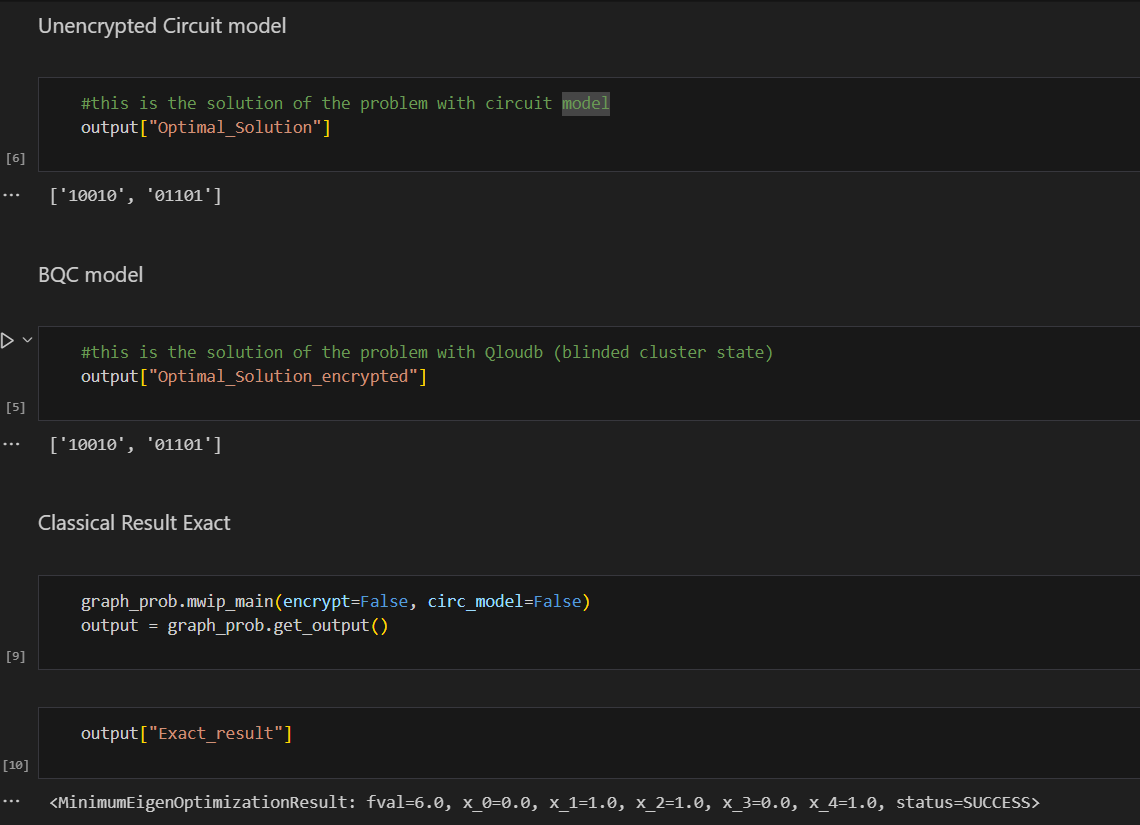} 
        \caption{Result (solution) of the problem solved by delegating the blinded circuit for the corresponding problem (represented by the command `Optimal\_Solution\_encrypted`) in the code.}
        \label{mwip_result}
    \end{subfigure}
    \caption{This is an example implementation of delegated MWIP problem using the given methodology. The Fig.\ref{mwip_result} shows the results which were obtained by delegating the `BQC model` to IBMQ server and running the `classical` and `circuit model` solution at the local system. Upon comparison the results from all the three executions match.}
    \label{MWIP}
\end{figure*}

We implemented our proposed methodology for the examples of bell state (see Fig. \ref{bell_example}), GHZ state preparations (see Fig. \ref{GHZ}) and a simple graph of Maximum Weighted Independent set Problem (MWIP) (see Fig. \ref{MWIP}) all of which were executed on IBMQ server using the QASM simulator backend. The blinded quantum circuits for the later two examples are not shown in results due to the circuit size being too large. Since all these examples were delegated to an IBMQ simulator with no error models, the results obtained match the ideal results. The dependency packages were qiskit-terra 0.24.1,
qiskit-aer 0.12.1, qiskit-ignis, qiskit-ibmq-provider 0.20.0, qiskit 0.43.2, qiskit-nature 0.5.2, qiskit-finance, qiskit-optimization 0.5.0, qiskit-machine-learning.

\section{Conclusion}
In conclusion, as quantum computing continues to advance, ensuring secure and private computation remains crucial for its practical adoption. The advancements we propose—modifying UBQC to remove the need for interactive measurements and incorporating the 8-state RSP protocol—address a critical gap between theoretical models and the limitations of current quantum platforms. While this approach does come with trade-offs, such as increased circuit depth and higher computational costs, and reduces the security from information-theoretical security to weaker game-based security, it enables fully classical clients to securely interact with quantum servers. Our method enhances the practicality of UBQC in real-world applications and paves the way for greater accessibility to quantum technology in current architecture. We also mentioned a drawback that an additional randomization parameter $r$ has to be set to 0 in our proposal, discussed in Sec.\ref{Step 1}. Addressing this additional constraint can be the subject of future work.

\section{Acknowledgements}
We gratefully acknowledge the use of IBM Quantum services and Qulabs Software (India) Pvt. Ltd. for this work. The views expressed in this paper are solely those of the authors and do not represent the official policies or positions of IBM or the IBM Quantum team.

The authors extend their heartfelt gratitude to Rikteem Bhowmick, Devesh Kumar, Keshav Singh, Devendra Mishra, Siddhi Ani Mali, Prof. Goutam Paul, Jay Shah, Omkar Bihani, Nilesh, and Prof. Kaushik Nandi for their invaluable assistance with coding, insightful discussions on filtering ideas, and continuous support throughout the project.

Finally, we express our deepest appreciation to our families for their unwavering love, encouragement, and support, which has been a cornerstone of our work.

\clearpage
\newpage
\bibliographystyle{apsrev4-2}
\bibliography{bibliography}
\section{Appendix}
\subsection{Equivalence circuit model to MBQC}
\label{cir2MBQC}
\begin{figure}[htb]
    \centering
    \begin{subfigure}[b]{0.5\textwidth}
        \centering
        \includegraphics[scale=0.7]{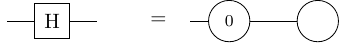}
    \end{subfigure}
    \begin{subfigure}[b]{0.5\textwidth}
        \centering
        \includegraphics[scale=0.7]{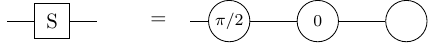}
    \end{subfigure}
     \begin{subfigure}[b]{0.5\textwidth}
        \centering
        \includegraphics[scale=0.7]{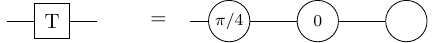}
    \end{subfigure}
     \begin{subfigure}[b]{0.5\textwidth}
        \centering
        \includegraphics[scale=0.7]{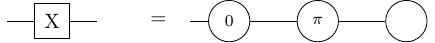}
    \end{subfigure}
    \caption{MBQC equivalent of universal single qubit gates}
    \label{mbqc_equi_1qubit}
\end{figure}

\begin{figure}[h]
    \centering
    \includegraphics[width=0.9\linewidth]{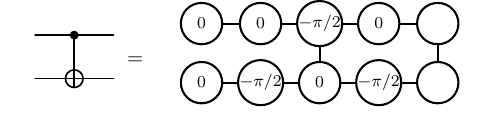}
    \caption{MBQC equivalent of CNOT gate}
    \label{mbqc_equi_2qubit}
\end{figure}
\subsection{Gate Equivalences}
\label{gate_identity}
\begin{itemize}
    \item $HZ_\beta H = X_\beta$
    \item $HX_\beta H = Z_\beta$
    \item $CZ_{(i,j)}Z_{i} = Z_{i}CZ_(i,j)$
    \item $CZ_{(i,j)}X_{i} = X_iZ_{j}CZ_(i,j)$
    \item $CZ_{(i,j)}Rz_{i}(\theta) = Rz_{i}(\theta)CZ_(i,j)$
\end{itemize}
\subsection{Non-interactive blindness}
\label{blindness_appendix}
We here show a non-interactive UBQC computation method for a simple but generalizable case given in Fig. \ref{1Dcircuit}. A $2D$ case follows the same computation with additional $Controlled-Z$ gate(s) in the circuit.
In the circuit, just before the measurement operation, the output state is given by:
\begin{align}
\label{eq_correction}
    \ket{\psi} = \frac{1}{2}[&\ket{00}(\ket{+}+e^{i\phi_1}\ket{+}+e^{i\phi_2}\ket{-}-e^{i(\phi_1+\phi_2)}\ket{-})
     +\nonumber \\ & \ket{01}(\ket{+}+e^{i\phi_1}\ket{+}-e^{i\phi_2}\ket{-}+e^{i(\phi_1+\phi_2)}\ket{-})+ \nonumber \\ & \ket{10}(\ket{+}-e^{i\phi_1}\ket{+}+e^{i\phi_2}\ket{-}+e^{i(\phi_1+\phi_2)}\ket{-})+ \nonumber \\ & \ket{11}(\ket{+}-e^{i\phi_1}\ket{+}-e^{i\phi_2}\ket{-}-e^{i(\phi_1+\phi_2)}\ket{-})]  \nonumber \\
     \text{or, } \ket{\psi}  = \frac{1}{2}[&\ket{00}(HRz(\phi_2)HRz(\phi_1))\ket{+}+ \nonumber \\ & \ket{01}X(HRz(\phi_2)HRz(\phi_1))\ket{+}+ \nonumber \\ &\ket{10}Z(HRz(\phi_2)HRz(\phi_1))\ket{+}+ \nonumber \\ &\ket{11}XZ(HRz(\phi_2)HRz(\phi_1))\ket{+}] \nonumber \\
     \text{or, } \ket{\psi} =& X^{m_2}Z^{m_1}(HRz(\phi_2)HRz(\phi_1))\ket{+} 
\end{align}
Since, $ZH = HX$ and $XRz(\phi) = Rz(\phi')$, where $\phi' = (-1)^{m_{1}^X}\phi+m_{1}^{Z}\pi$, we can rewrite above as $\ket{\psi} =  X^{m_2}HX^{m_1}Rz(\phi_2)HRz(\phi_1)\ket{+} =  X^{m_2}HRz(\phi')X^{m_1}HRz(\phi_1))\ket{+}$, which is same the first part of interactive correction in \cite{Broadbent_2009} . Till now, we have shown that all the corrections can be pushed to the end of the circuit, but, we need to show the same for Fig. \ref{1Dcircuit_blind} which is the case for UBQC.\\
We will now show that by introducing the blinding parameters, $\ket{+} \rightarrow \ket{+_{\theta_{i}}}$ for input qubits and $\phi_i \rightarrow \delta_i$ we can still get the same implementation result as Eq.\ref{eq_correction}.
\begin{align*}
    \ket{\psi'_0} = \frac{1}{2}[(\ket{0}+e^{i\theta_1}\ket{1})(\ket{0}+e^{i\theta_2}\ket{1})]\ket{+} 
\end{align*}
\begin{widetext}
\begin{multline}
    \label{eq_blinding}
    \ket{\psi'_1} = \frac{1}{2\sqrt{2}}(\ket{000}+\ket{001}+e^{i\theta_2}\ket{010}-e^{i\theta_2}\ket{011}+e^{i\theta_1}\ket{100}+e^{i\theta_1}\ket{101} -e^{i(\theta_1+\theta_2)}\ket{110}+e^{i(\theta_1+\theta_2)}\ket{111})\\
\ket{\psi'_2} = \frac{1}{2\sqrt{2}}(\ket{000}+\ket{001}+e^{i(\delta_2+\theta_2)}\ket{010}-e^{i(\delta_2+\theta_2)}\ket{011}+
e^{i(\delta_1+\theta_1)}\ket{100}+e^{i(\delta_1+\theta_1)}\ket{101} -e^{i(\delta_1+\theta_1+\delta_2+\theta_2)}\ket{110}+\\e^{i(\delta_1+\theta_1+\delta_2+\theta_2)}\ket{111})\\
\ket{\psi'_3} = \frac{1}{2\sqrt{2}}[(\ket{0}+\ket{1})(\ket{0}+\ket{1})\ket{0}+(\ket{0}+\ket{1})(\ket{0}+\ket{1})\ket{1}+e^{i(\delta_2+\theta_2)}(\ket{0}+\ket{1})(\ket{0}-\ket{1})\ket{0}-e^{i(\delta_2+\theta_2)}(\ket{0}+\ket{1})(\ket{0}-\ket{1})\ket{1}+\\e^{i(\delta_1+\theta_1)}(\ket{0}-\ket{1})(\ket{0}+\ket{1})\ket{0}+e^{i(\delta_1+\theta_1)}(\ket{0}-\ket{1})(\ket{0}+\ket{1})\ket{1}-e^{i(\delta_1+\theta_1+\delta_2+\theta_2)}(\ket{0}-\ket{1})(\ket{0}-\ket{1})\ket{0}+e^{i(\delta_1+\theta_1+\delta_2+\theta_2)}\\(\ket{0}-\ket{1})(\ket{0}-\ket{1})\ket{1}]\\
\ket{\psi'} = \ket{00}[\ket{+}+e^{i(\delta_1+\theta_1)}\ket{+}+e^{i(\delta_2+\theta_2)}\ket{-}+e^{i(\delta_1+\theta_1+\delta_2+\theta_2)}\ket{-} +
\ket{01}[\ket{+}+e^{i(\delta_1+\theta_1)}\ket{+}-e^{i(\delta_2+\theta_2)}\ket{-}+\\e^{i(\delta_1+\theta_1+\delta_2+\theta_2)}\ket{-}+\ket{10}[\ket{+}-e^{i(\delta_1+\theta_1)}\ket{+}+e^{i(\delta_2+\theta_2)}\ket{-}-e^{i(\delta_1+\theta_1+\delta_2+\theta_2)}\ket{-}]
     \end{multline}
     \end{widetext}

     putting $\delta_i = \phi_i-\theta_i$,
\begin{align}
\label{eq_blinding2}
    \ket{\psi'} = \frac{1}{2}[&\ket{00}(\ket{+}+e^{i\phi_1}\ket{+}+e^{i\phi_2}\ket{-}-e^{i(\phi_1+\phi_2)}\ket{-})
     +\nonumber \\ & \ket{01}(\ket{+}+e^{i\phi_1}\ket{+}-e^{i\phi_2}\ket{-}+e^{i(\phi_1+\phi_2)}\ket{-})+ \nonumber \\ & \ket{10}(\ket{+}-e^{i\phi_1}\ket{+}+e^{i\phi_2}\ket{-}+e^{i\phi_1+\phi_2}\ket{-})+ \nonumber \\ & \ket{11}(\ket{+}-e^{i\phi_1}\ket{+}-e^{i\phi_2}\ket{-}-e^{i(\phi_1+\phi_2)}\ket{-})]  \nonumber \\
     \end{align}
We will now show that why an additional parameter $r_{i,j}$
doesn't work in the non-interactive system. If we put $\delta_i = \phi_i-\theta_i+r_i\pi$ in Eq. \ref{eq_blinding}, then,  
\begin{widetext}
\begin{align}
\label{eq_blinding3}
    \ket{\psi'} = \frac{1}{2}[&\ket{00}(\ket{+}+e^{i(\phi_1+r_1\pi)}\ket{+}+e^{i(\phi_2+r_2\pi)}\ket{-}-e^{i(\phi_1+\phi_2+r_1\pi+r_2\pi)}\ket{-})
     +\ket{01}(\ket{+}+e^{i(\phi_1+r_1\pi)}\ket{+}-\nonumber \\& e^{i(\phi_2+r_2\pi)}\ket{-}+  e^{i(\phi_1+\phi_2+r_1\pi+r_2\pi)}\ket{-})+ \ket{10}(\ket{+}-e^{i(\phi_1+r_1\pi)}\ket{+}+e^{i(\phi_2+r_2\pi)}\ket{-}+\nonumber \\ & e^{i(\phi_1+\phi_2+r_1\pi+r_2\pi)}\ket{-})+  \ket{11}(\ket{+}-e^{i(\phi_1+r_1\pi)}\ket{+}-e^{i(\phi_2+r_2\pi)}\ket{-}-e^{i(\phi_1+\phi_2+r_1\pi+r_2\pi)}\ket{-})]  \nonumber \\
     \end{align}
     \end{widetext}
     At this stage, it can be very easy to see that for the case $r_1,r_2 = 0,0$ this equation is same as Eq.\ref{eq_blinding2}, but the analysis becomes a little trickier for other pairs of $r_1,r_2$, as now we are applying corrections to the classical output, $s$. Therefore, we will now resort to matching the coefficients with the coefficient, $c_1 = 1+e^{i\phi_1}+e^{i(\phi_2)}-e^{i(\phi_1+\phi_2)}$ for $s' = 0$ and $c_2 = 1+e^{i\phi_1}-e^{i(\phi_2)}+e^{i(\phi_1+\phi_2)}$ for $s'=1$, associated to the expected output bit $s'$  to see if our implementation is correct in the remaining cases, as these coefficients contribute to the output probability distribution. These coefficients come from the first term in Eq.\ref{eq_blinding2}, when the output state is written in the computational basis. \\
     \textbf{Case: $r_1 = 0, r_2 = 1, r_1 \oplus r_2 = 1 \implies s' = s \oplus 1$}.\\
     In this case, the coefficient term for first and second qubit measurement outcomes as $00$ becomes:\\
     \begin{align}
         (1+(-1)^{r_1}e^{i\phi_1}+(-1)^{r_2}e^{i\phi_2}-(-1)^{r_1 \oplus r_2}e^{i(\phi_1+\phi_2)})\ket{0}+\nonumber \\
         (1+(-1)^{r_1}e^{i\phi_1}-(-1)^{r_2}e^{i\phi_2}+(-1)^{r_1 \oplus r_2}e^{i(\phi_1+\phi_2)})\ket{1} \nonumber\\ 
          c_2\ket{0} + c_1\ket{1}
     \end{align}
     These flipped coefficients reflects in the probability distribution outcome which is corrected by bitflip operation.
     \textbf{Case: $r_1 = 1, r_2 = 0, r_1 \oplus r_2 = 1 \implies s' = s \oplus 1$}.\\
     Similar to above, the coefficient we get are 
     \begin{align}
         (1+(-1)^{r_1}e^{i\phi_1}+(-1)^{r_2}e^{i\phi_2}-(-1)^{r_1 \oplus r_2}e^{i(\phi_1+\phi_2)})\ket{0}+\nonumber \\
         (1+(-1)^{r_1}e^{i\phi_1}-(-1)^{r_2}e^{i\phi_2}+(-1)^{r_1 \oplus r_2}e^{i(\phi_1+\phi_2)})\ket{1} \nonumber\\ 
     \end{align}
     These coefficients doesn't correspond to either $c_1$ or $c_2$, hence bitflip correction doesn't work and might be the future scope of this work.
\end{document}